\documentclass[twocolumn,prb,10pt,aps,superscriptaddress,floatfix]{revtex4-1}
\usepackage[english]{babel}
\usepackage{graphicx}
\usepackage[pdftex]{epsfig}
\usepackage{ragged2e}
\usepackage{amsmath,amssymb}
\usepackage{times}
\usepackage[colorlinks,linkcolor=blue,citecolor=blue,urlcolor=blue]{hyperref}
\usepackage{color}
\usepackage[table,xcdraw]{xcolor}
\usepackage{epstopdf}
\usepackage{physics}
\usepackage[export]{adjustbox}
\usepackage{dsfont}

\newcommand{\be}{\begin{equation}}
\newcommand{\ee}{\end{equation}}

\newcommand{\bee}{\begin{equation*}}
\newcommand{\eee}{\end{equation*}}

\newcommand{\cacro}{Ca$_{10}$Cr$_7$O$_{28}$}

\begin{document}
\title{Signatures for spinons in the quantum spin liquid candidate Ca$_{10}$Cr$_7$O$_{28}$}
\author{Jonas Sonnenschein}
\affiliation{Dahlem Center for Complex Quantum Systems and Institut f\"ur Theoretische Physik, Freie Universit\"{a}t Berlin, Arnimallee 14, 14195 Berlin, Germany}
\affiliation{Helmholtz-Zentrum f\"{u}r Materialien und Energie, Hahn-Meitner-Platz 1, 14109 Berlin, Germany}
\author{Christian Balz}
\affiliation{Neutron Scattering Division, Oak Ridge National Laboratory, Oak Ridge, TN 37831, U.S.A.}
\author{Ulrich Tutsch}
\author{Michael Lang}
\affiliation{Physikalisches Institut, Goethe-Universit\"at Frankfurt, Max-von-Laue-Strasse 1, 60438 Frankfurt, Germany}
\author{Hanjo Ryll}
\affiliation{Helmholtz-Zentrum f\"{u}r Materialien und Energie, Hahn-Meitner-Platz 1, 14109 Berlin, Germany}
\author{Jose A. Rodriguez-Rivera}
\affiliation{NIST Center for Neutron Research, National Institute of Standards and Technology, 20899 Gaithersburg, USA}
\affiliation{Department of Materials Science, University of Maryland, College Park, 20742 Maryland, USA}
\author{A. T. M. Nazmul Islam}
\affiliation{Helmholtz-Zentrum f\"{u}r Materialien und Energie, Hahn-Meitner-Platz 1, 14109 Berlin, Germany}
\author{Bella Lake}
\affiliation{Helmholtz-Zentrum f\"{u}r Materialien und Energie, Hahn-Meitner-Platz 1, 14109 Berlin, Germany}
\affiliation{Institut f\"ur Festk\"orperphysik, Technische Universit\"at Berlin, 10623 Berlin, Germany}
\author{Johannes Reuther}
\affiliation{Dahlem Center for Complex Quantum Systems and Institut f\"ur Theoretische Physik, Freie Universit\"{a}t Berlin, Arnimallee 14, 14195 Berlin, Germany}
\affiliation{Helmholtz-Zentrum f\"{u}r Materialien und Energie, Hahn-Meitner-Platz 1, 14109 Berlin, Germany}
\date{\today}

\begin{abstract}
We present new experimental low-temperature heat capacity and detailed dynamical spin-structure factor data for the quantum spin liquid candidate material Ca$_{10}$Cr$_7$O$_{28}$. The measured heat capacity shows an almost perfect linear temperature dependence in the range $0.1$ K $\lesssim T\lesssim0.5$ K, reminiscent of fermionic spinon degrees of freedom. The spin structure factor exhibits two energy regimes of strong signal which display rather different but solely diffuse scattering features. We theoretically describe these findings by an effective spinon hopping model which crucially relies on the existence of strong ferromagnetically coupled triangles in the system. Our spinon theory is shown to naturally reproduce the overall weight distribution of the measured spin structure factor. Particularly, we argue that various different observed characteristic properties of the spin structure factor and the heat capacity consistently indicate the existence of a spinon Fermi surface. A closer analysis of the heat capacity at the lowest accessible temperatures hints towards the presence of weak $f$-wave spinon pairing terms inducing a small partial gap along the Fermi surface (except for discrete nodal Dirac points) and suggesting an overall $\mathds{Z}_2$ quantum spin liquid scenario for Ca$_{10}$Cr$_7$O$_{28}$.
\end{abstract}
\maketitle

\section{Introduction}
The hunt for experimental realizations of novel topological quantum states is one of the most thriving research themes in modern condensed matter physics. Quantum spin liquids attract special interest since they combine various exotic phenomena which completely fall outside the traditional Landau paradigm of symmetry broken phases of matter.\cite{anderson73,balents10,savary17} Instead of the classical long-range order of more conventional magnets, quantum spin liquids exhibit topological order which cannot be described by any local order parameter.\cite{wen90} Similarly, as a fundamental difference to the well-known spin waves (or magnons) of magnetically ordered states, quantum spin liquids harbor fractional spin excitations which carry anyonic quasiparticle statistics. Particularly, the fundamental spinful quasiparticles of a quantum spin liquid are so-called spinons which can be thought of as a fraction (i.e., half) of a conventional $\Delta S=1$ spin flip excitation.\cite{lake05,mourigal13,han12}
 
The experimental and theoretical investigation of these phenomena, however, poses significant challenges. For example, due to its long-range entangled nature, there are currently no experimental techniques available which can directly identify topological order. Concerning the excitations of a quantum spin liquid, their fractional nature prohibits the creation of single spinons, however, neutron scattering at least allows their two-particle continuum to be probed which typically forms broad and diffuse patterns in the spin structure factor. Such diffuse neutron scattering has indeed been observed in various promising spin liquid candidate materials such as the kagome system Herbertsmithite\cite{helton07,han12,han16} [ZnCu$_3$(OH)$_6$Cl$_2$], the triangular magnet YbMgGaO$_4$,\cite{shen16,paddison16,shen18} and the Kitaev honeycomb material $\alpha$-RuCl$_3$\cite{banerjee16,banerjee17,do17,banerjee18,balz19} (even though the latter material is known to order at low temperatures\cite{sears15}). Whether or not these measured responses are indeed signatures of spinons is not yet completely settled since it is well known that non-fractional phenomena such as multi-magnon continua, spin-glass behavior or chemical disorder effects may also give rise to broad excitation spectra.\cite{dally14,zhu17,li17,ma18,parker18,kimchi18} As a further complication from the theory side, except for special coupling scenarios as realized in the exactly solvable Kitaev model,\cite{kitaev06} it is extremely challenging to calculate the spin excitation spectrum starting from a generic spin Hamiltonian. For this reason, it is often easier to theoretically investigate quantum spin liquids based on an effective model for its fractional excitations.

Apart from the information drawn from neutron experiments, thermodynamic properties can help unraveling quantum spin liquid behavior. According to current understanding, spinons may behave like chargeless fermions,\cite{hao09,hao13,normand16} i.e., similar to electrons in a metal they can lead to a heat capacity and thermal conductivity linear in temperature. Indeed, approximate linearity of these quantities has been observed in various spin liquid candidate materials.\cite{yamashita08,yamashita10,yamashita11} Furthermore, thermal Hall conductivity measurements have recently shown promising signatures of fractional edge states in the Kitaev candidate compound $\alpha$-RuCl$_3$.\cite{kasahara18}

\begin{figure}[t]
\centering
\includegraphics[width=0.8\linewidth]{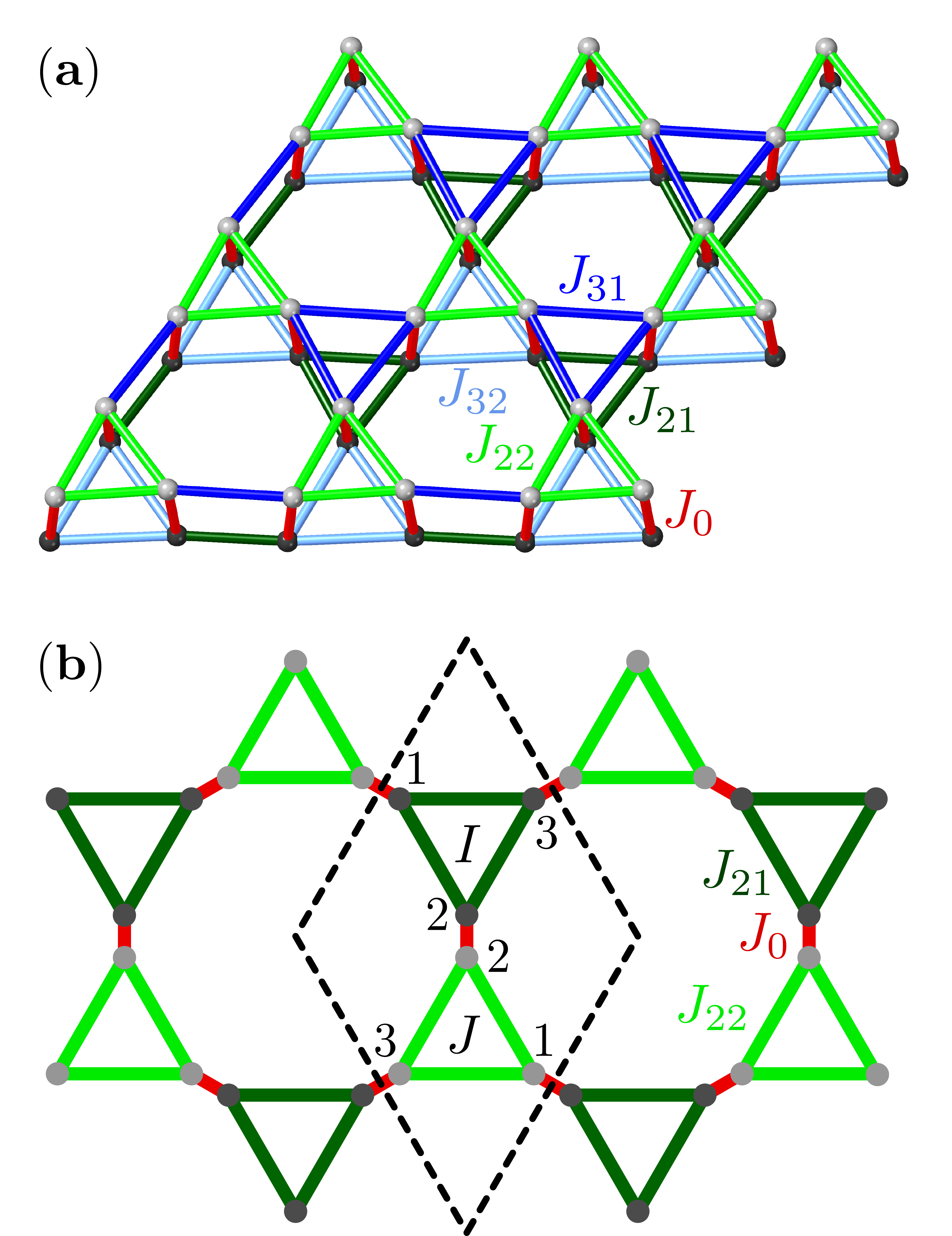}
\caption{(a) Bilayer kagome lattice as realized in Ca$_{10}$Cr$_7$O$_{28}$. The differently colored bonds carry the interactions $J_0$, $J_{21}$, $J_{22}$, $J_{31}$, and $J_{32}$ as indicated in the figure, see also Table~\ref{tab}. (b) Effective decorated honeycomb lattice arising from a projection of the ferromagnetically coupled triangles (green triangles labelled $I$, $J$) of the bilayer kagome system into one plane. Bonds are colored and labelled in the same way as in (a), except the antiferromagnetic (blue) bonds which are not shown. Note that sites coupled by the vertical ferromagnetic interlayer couplings (red lines) almost coincide in their position after projection. We have, hence, increased their in-plane distance in this illustration for better visibility. Dark gray (light gray) dots denote sites in the lower (upper) plane. Dashed lines mark the boundaries of the unit cell and numbers label the sites within ferromagnetically coupled triangles.}
\label{fig:bilayer}
\end{figure}

In this paper, we address the aforementioned opportunities and challenges in identifying quantum spin liquids based on the compound Ca$_{10}$Cr$_7$O$_{28}$.\cite{balz16,balz17,balz17_2,pohle17,biswas18,augustine19} Particularly, we demonstrate that new heat capacity and single crystal neutron scattering data in conjunction with an effective model for the low-energy excitations allows an interpretation in terms of emergent spinon degrees of freedom that is remarkably straightforward and consistent. In a previous publication by some of the authors, Ca$_{10}$Cr$_7$O$_{28}$ has been shown to feature striking properties of a quantum spin liquid such as an absence of magnetic long-range order down to at least 19 mK, persistent spin dynamics at low temperatures and a diffuse scattering signal in neutron experiments.\cite{balz16} Furthermore, in contrast to other spin-liquid candidate systems, Ca$_{10}$Cr$_7$O$_{28}$ is characterized by a larger immunity to chemical disorder since site mixing is suppressed by distinctly different ionic radii. The strong quantum fluctuations in this compound can be explained by a special frustration mechanism arising due to interacting spin-$1/2$ magnetic moments from Cr$^{5+}$ ions, arranged in a stacked bilayer kagome geometry, see Fig.~\ref{fig:bilayer}. Based on neutron scattering in a magnetic field combined with a spin wave analysis, a microscopic Heisenberg Hamiltonian $H=\frac{1}{2}\sum_{i,j}\mathcal{J}_{ij}{\mathbf S}_i{\mathbf S}_j$ has been identified which features five different interactions $\mathcal{J}_{ij}$ on geometrically distinct bonds of the bilayer system denoted by $J_0$, $J_{21}$, $J_{22}$, $J_{31}$, and $J_{32}$ (see Table~\ref{tab} for the coupling strengths found in Refs.~\onlinecite{balz16,balz17}).

In contrast to the ideal antiferromagnetic kagome Heisenberg model discussed in the context of Herbertsmithite, the magnetic lattice of Ca$_{10}$Cr$_7$O$_{28}$ exhibits various peculiarities. Particularly, the `up' and `down' triangles of both kagome layers are all symmetry-inequivalent resulting in four differently coupled equilateral triangles carrying antiferromagnetic {\it and} ferromagnetic interactions $J_{21}$, $J_{22}$, $J_{31}$, $J_{32}$ where the ferromagnetic couplings ($J_{21}$ and $J_{22}$) are even the largest ones. 
Ferromagnetic (green) and antiferromagnetic (blue) triangles alternate within each layer and the two layers are stacked so that the ferromagnetic triangles in the first layer lie on top of the antiferromagnetic triangles in the second layer and vice versa. The fifth coupling $J_0$ is also ferromagnetic and vertically connects sites in the two layers. While the two layers individually only exhibit relatively weak spin frustration (three spins on ferromagnetic triangles approximately combine into spins $3/2$ living on an antiferromagnetic triangular lattice in each layer\cite{pohle17,biswas18}) the interlayer interactions $J_0$ induce very strong frustration effects and are primarily responsible for the destruction of magnetic order. Despite the complexity of the system, numerical studies confirmed the non-magnetic ground state of the proposed spin Hamiltonian\cite{balz16,pohle17,biswas18,augustine19} and reproduced the overall weight distribution of the static spin structure factor.\cite{balz16,pohle17,biswas18} Yet, a coherent interpretation and explanation of the measured observables in terms of emergent spinon quasiparticles as will be presented below has not been developed so far.
\begin{table}[t]
\centering
\begin{tabular}{ccccc}
 \hline \hline
$J_0$&$J_{21}$&$J_{22}$&$J_{31}$&$J_{32}$\\
$-0.08(4)$&$-0.76(5)$&$-0.27(3)$&$0.09(2)$&$0.11(3)$\\
\hline\hline
\end{tabular}
\caption{Exchange couplings of Ca$_{10}$Cr$_7$O$_{28}$ as determined in Refs.~\onlinecite{balz16,balz17}. All couplings are given in meV.}
\label{tab}
\end{table}

The effective spinon model which we propose in this work relies on new experimental heat capacity and dynamical spin structure factor data on Ca$_{10}$Cr$_7$O$_{28}$. The presented heat capacity data ranges down to lower temperatures as compared to our previous work (37 mK versus 330 mK in Ref.~\onlinecite{balz16}). Interestingly, the newly resolved temperature regime shows a nearly perfect linear behavior over almost one order of magnitude in $T$, reminiscent of fermionic spinons. Our single crystal neutron scattering data captures the dynamical spin structure with better energy resolution and, hence, shows various additional details which previously remained unresolved. We find that the excitation spectrum reveals two clearly separated diffuse bands of strong response where the one at lower energies exhibits a characteristic V-shape in energy around the $\Gamma$-point. The central assumption in the ensuing theoretical analysis amounts to attributing the absence of magnetic long-range order, the linear heat capacity and the overall diffuse magnetic scattering to the existence of a quantum spin liquid ground state with emergent spinon excitations. Various known properties of Ca$_{10}$Cr$_7$O$_{28}$ such as its spin-isotropy and the presence of strong ferromagnetic interactions allow us to put constraints on the low-energy dynamics of spinons. This naturally leads us to an effective spinon hopping model which reproduces key experimental features even without a fine-tuning of parameters. Particularly, we find that the weight distribution in the two bands of scattering is rooted in the special pattern of ferromagnetic bonds. Similarly, the linear heat capacity and V-shaped spin structure factor are explained by an (approximate) spinon Fermi surface where a small deviation from linearity at the lowest temperatures possibly indicates the formation of a partial gap along the Fermi level (except for discrete gapless Dirac points) due to weak spinon pairing. We, therefore, conclude that a gapless $\mathds{Z}_2$ spin liquid is the most plausible scenario for Ca$_{10}$Cr$_7$O$_{28}$.

The paper is organized as follows: In Sec.~\ref{experiment} we provide experimental details about the measurements performed (Sec.~\ref{exp_details}) and present the new low temperature heat capacity and detailed dynamical spin structure factor data (Sec.~\ref{exp_results}). The theoretical analysis in Sec.~\ref{theory} starts by reviewing the general parton theory for quantum spin liquids (Sec.~\ref{partons}) before the microscopic spinon model for Ca$_{10}$Cr$_7$O$_{28}$ is addressed more specifically (Sec.~\ref{cacro}). The overall implications of this model for the spin structure factor are discussed in Sec.~\ref{general_weight_dist} and a more detailed comparison with the experimental data can be found in Sec.~\ref{comparison}. In Sec.~\ref{low_e} we also include the heat capacity into our analysis which modifies the spinon model at low energies. The papers ends with a conclusion in Sec.~\ref{conclusion}.

\section{Measured spin-structure factor and heat capacity}\label{experiment}
\subsection{Experimental Details}\label{exp_details}

Single crystal samples of {\cacro} were prepared according to the procedure described in Ref.~\onlinecite{balz17_2}. The heat capacity was measured on two different calorimeters. The first measurement was performed on a $0.93$~mg single crystal in the temperature range $0.3-6.5$~K using a quasi-adiabatic relaxation method in combination with a $^3$He cryostat at the Core Lab for Quantum Materials, Helmholtz-Zentrum Berlin. The second measurement was performed between $37$~mK and $1.7$~K on a larger $11.1$~mg single crystal at the Physikalisches Institut, Goethe-Universit\"at Frankfurt using a home-made calorimeter operated in both a relaxation mode as well as a continuous heating mode.

Inelastic neutron scattering was measured on the MACS II spectrometer (NIST Center for Neutron Research, Gaithersburg, USA). Two co-aligned rod-shaped single crystals with a total mass of $1.7$~g and a mosaicity of less than $2^\circ$ were used. The kagome bilayer $[H,K,0]$ plane was aligned with the horizontal scattering plane and the temperature was kept below $0.1$~K throughout the measurement. The final energy was fixed to $2.5$~meV for energy transfers $E\leq0.25$~meV and $3.0$~meV for energy transfers $E\geq0.25$~meV. An empty sample holder measurement was used for background subtraction. The energy resolution broadening of our data increases with incident energy from $\Delta E=0.1$ meV at $E=0.15$ meV to $\Delta E=0.3$ meV at $1.55$ meV. The data is visualized using the DAVE software package \cite{azuah09}.
\begin{figure}[t]
\centering
\includegraphics[width=0.8\linewidth]{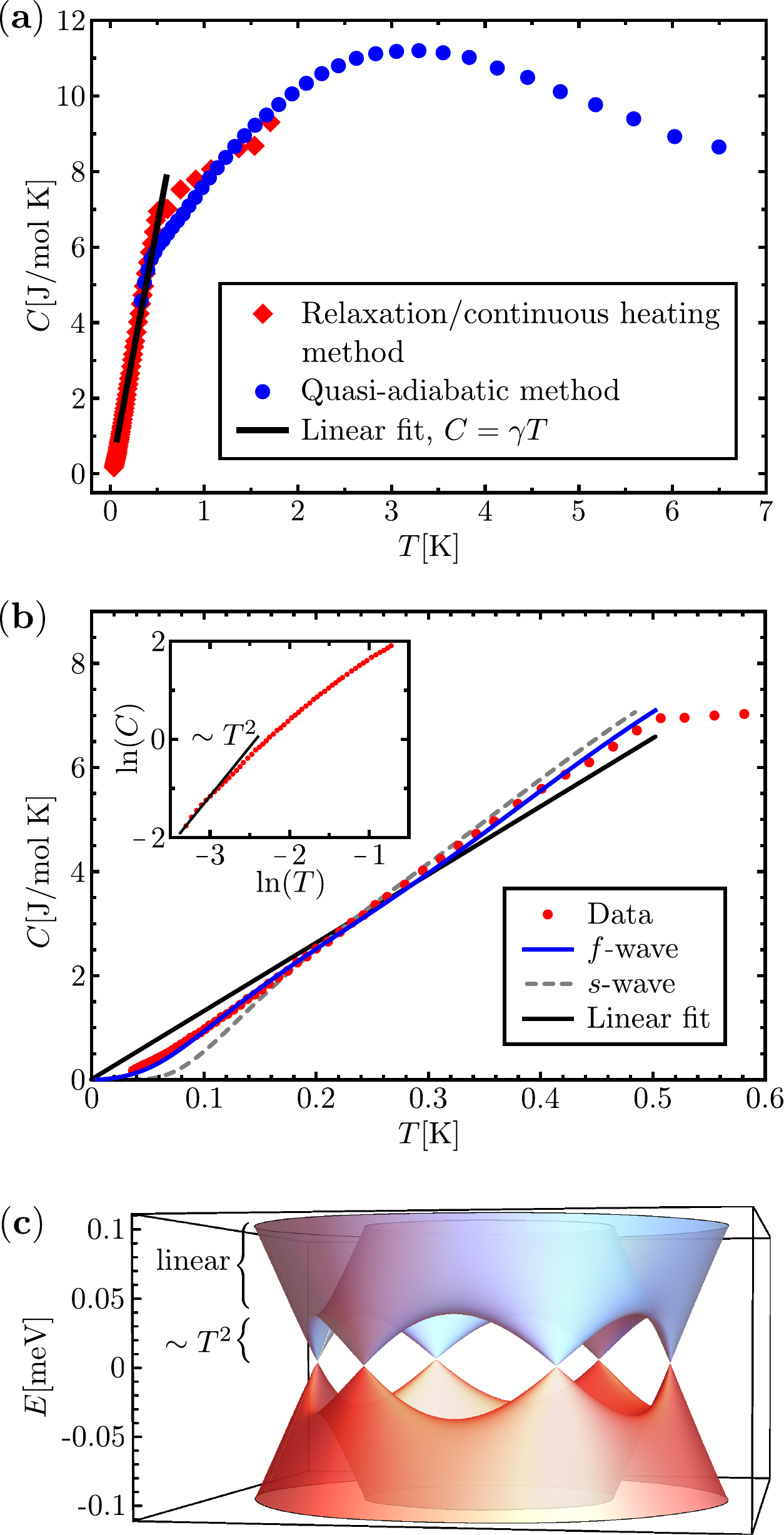}
\caption{(a) Measured heat capacity of Ca$_{10}$Cr$_7$O$_{28}$ in an extended temperature range. Shown are two different data sets as described in Sec.~\ref{exp_details} . The black line indicates the approximate linear behavior at low temperatures. (b) Enlarged view of the low-temperature behavior of the measured heat capacity (relaxation/continuous heating method). Several fits are shown: Linear temperature dependence as obtained for an intact Fermi surface (black line), $s$-wave pairing model with a ${\mathbf k}$-independent gap $\Delta_s=0.039$ meV (gray dashed line), and $f$-wave pairing model with the gap function in Eq.~(\ref{f_wave}) using $\Delta_f=0.039$ meV (blue line). Inset: Heat-capacity data in a double-logarithmic plot. For comparison, the black line shows a $\sim T^2$ temperature dependence. (c) Low-energy spinon band structure for the $f$-wave pairing model in Eq.~(\ref{f_wave}) with $\Delta_f=0.039$ meV. The energy regimes which lead to a linear and quadratic temperature dependence of the heat capacity are indicated.}
\label{fig:heat_capacity}
\end{figure}

\subsection{Experimental Results}\label{exp_results}

We start discussing the heat capacity data of {\cacro} which is shown in Fig.~\ref{fig:heat_capacity}(a) up to $7$~K. Note that the figure contains two independent data sets, as described above, where the one at higher temperatures has already been presented in Ref.~\onlinecite{balz16}. In the temperature range $0$ - $7$ K the phonon contribution was calculated to be negligible and the total specific heat was found to be of magnetic origin \cite{balz17}. The broad peak centered at $T\approx3$~K indicates the onset of short-ranged magnetic correlations. Furthermore, both data sets show a small kink at $500$~mK. This feature is consistent with a crossover in the same temperature regime observed in muon spin resonance measurements where the spin fluctuation rate becomes constant and the system enters a low-temperature phase of persistent slow dynamics\cite{balz16}. We, therefore, interpret this kink as crossover into a quantum spin liquid phase. The two data sets differ somewhat in the sharpness of the kink at $500$~mK which indicates a small sample dependence of this feature. Most importantly, no lambda-like anomaly indicative of long-range magnetic order is observed down to the lowest measured temperature of $37$~mK. Remarkably, the low-temperature data (and also the lowest data points of the high temperature measurement) exhibit an intriguing almost perfect linear behavior of the heat capacity below $500$~mK. An enlarged view of the heat capacity in this temperature regime [Fig.~\ref{fig:heat_capacity}(b)] reveals that the linear behavior persists down to approximately $100$~mK and shows a slight decrease of the slope for lower temperatures. This linear dependence is reminiscent of fermionic spinon quasiparticles and will be discussed in more detail in Sec.~\ref{low_e}.

\begin{figure}[t]
\centering
\includegraphics[width=0.99\linewidth]{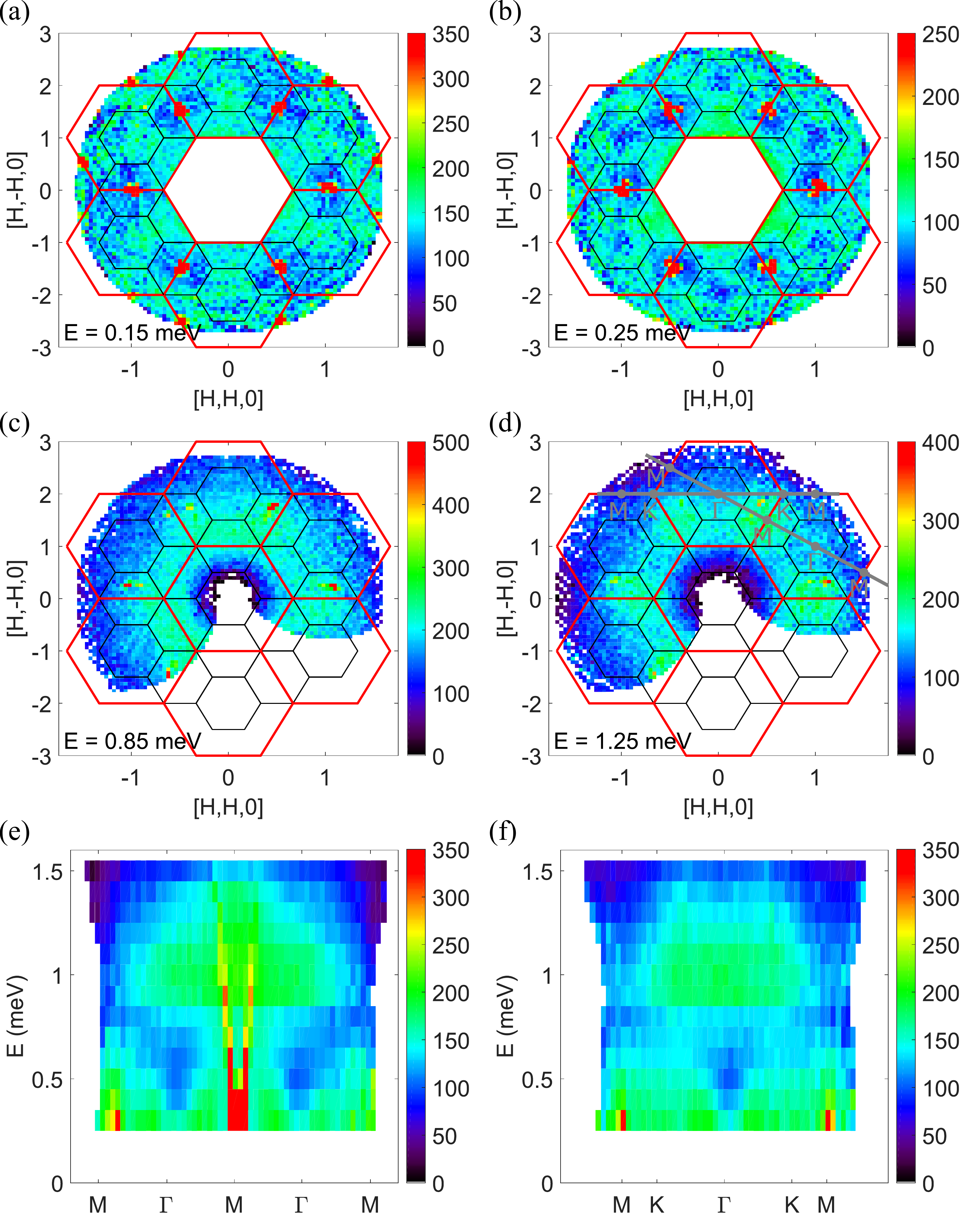}
\caption{Inelastic neutron scattering data of \cacro. (a) - (d) Constant energy slices as a function of the momentum transfer in the kagome bilayer plane. The black (red) hexagons indicate the boundaries of the first (extended) Brillouin zone. The energy transfer is indicated in each plot. (e) - (f) Energy versus momentum slices along two high-symmetry directions within the kagome bilayer plane. The two momentum cuts are illustrated by the gray lines in (d). Note that the color scale is different in each subfigure. The sharp features appearing in red outside the color scale are phonons dispersing out of nuclear Bragg peaks. Note that the constant energy slices in (a) and (b) were measured with a final energy of $E_f=2.5$ meV which leads to an overall lower intensity compared to the constant energy slices in (c) and (d) measured with $E_f=3$ meV. For the energy versus momentum slices in (e) and (f) all data was taken with $E_f=3$ meV. Furthermore, the data in these two plots was collected by integrating the signal over $\pm0.2$ r.l.u. in directions perpendicular to the respective cuts.}
\label{fig:neutron}
\end{figure}

Next, we present our new single crystal inelastic neutron scattering data which we display as selected cuts in momentum and energy space. While Fig.~\ref{fig:neutron}(a) - (d) show the scattering signal for several fixed energies $E$, in Fig.~\ref{fig:neutron}(e), (f) we display slices as a function of energy along two different momentum space directions [see Fig.~\ref{fig:neutron}(d) where these momentum cuts are indicated by the gray lines]. Two different contributions to the scattering cross section are immediately evident. The sharp intense features (appearing in red on our color scale) are acoustic phonons dispersing out of strong nuclear Bragg peaks. Note that the phonon scattering illustrates the sharpness of the ${\mathbf q}$-resolution. The broad diffuse signal of intermediate intensity (green on our color scale) has a $|\mathbf{q}|$ dependence consistent with the Cr$^{5+}$ magnetic form factor and hence measures the magnetic scattering cross section of {\cacro}. The latter is directly proportional to the dynamical spin-structure factor $\mathcal{S}(\mathbf{q},E)$ which we will investigate theoretically in Sec.~\ref{theory}. 

The spin excitation spectrum $\mathcal{S}(\mathbf{q},E)$ exhibits several interesting features some of which were already described in Ref.~\onlinecite{balz16}. Firstly, the scattering signal is divided into two energy bands with a region of weak intensity separating them. The low-$E$ band [Fig.~\ref{fig:neutron}(a), (b)] extends to energy transfers up to $0.6$~meV while the high-$E$ band [Fig.~\ref{fig:neutron}(c), (d)] spreads between $0.8$ and $1.4$~meV. The separation of the response into these bands is most evident from the energy-momentum slices of Fig.~\ref{fig:neutron}(e), (f). We find sizable scattering intensities down to the lowest accessible energies ($0.15$~meV) indicating a gapless spin excitation spectrum. At these energies the intensities are centered around the $\Gamma$ point as seen in Fig.~\ref{fig:neutron}(a). Increasing the energy within the low-$E$ region the response disperses from the $\Gamma$ point to the boundary of the first Brillouin zone (black hexagons in Fig.~\ref{fig:neutron}). Within the experimental resolution this dispersion appears to be linear with energy which manifests in a V-shaped onset of intensity dispersing out of the $\Gamma$ point, see Fig.~\ref{fig:neutron}(e), (f). The magnetic response in the upper band is found to be fundamentally different with intensity concentrated around the boundary of the extended Brillouin zone (which is identical to the fourth Brillouin zone illustrated by the red hexagons in Fig.~\ref{fig:neutron}). Within the upper band the intensity varies but the overall shape of $\mathcal{S}(\mathbf{q},E)$ remains almost unchanged. Most importantly, neither of the two regimes shows sharp magnetic excitation modes as would be expected for conventional magnetically ordered states. In addition to the results in Fig.~\ref{fig:neutron} we show the full neutron scattering data for further fixed energies in the appendix.

\section{Theoretical modeling}\label{theory}
\subsection{General parton description of quantum spin liquids}\label{partons}
In this section, we develop and discuss a microscopic model for the magnetic excitations of Ca$_{10}$Cr$_7$O$_{28}$ which qualitatively reproduces the measured data presented in the previous section and, hence, allows for additional insights into the fundamental quasiparticles of this compound. Particularly, due to strong experimental evidence, we base the following considerations on the assumption that Ca$_{10}$Cr$_7$O$_{28}$ realizes a quantum spin-liquid ground state. We shall, therefore, interpret the linear heat capacity and the diffuse scattering in neutron experiments as signatures of spinons which represent the elementary {\it spinful} excitations in a quantum spin liquid.\cite{wen91,read91,balents10,savary17} Before we discuss Ca$_{10}$Cr$_7$O$_{28}$ more specifically, we first briefly review some general properties of spinon excitations in quantum spin liquids and explain how these quasiparticles can be theoretically modeled. 

Spinons can generally be viewed as half of a physical $\Delta S=1$ spin flip and are therefore referred to as {\it fractional} quasiparticles. This property is most conveniently expressed in a parton picture where spinons are modeled by spinful fermionic\cite{abrikosov65,arovas88,wen91,wen02} or bosonic\cite{arovas88,wang06,wang10,messio13} creation/annihilation operators $f_{i\alpha}^\dagger$, $f_{i\alpha}$ where $i$ is the site index and $\alpha=\uparrow,\downarrow$ labels the spin degree of freedom. The fractional property of spinons implies that the physical spin operator ${\mathbf S}_i$ becomes a composite object when expressed in terms of spinon operators. This is described by the relation
\begin{equation}
S_i^\mu=\frac{1}{2}\sum_{\alpha\beta}f_{i\alpha}^\dagger \sigma_{\alpha\beta}^\mu f_{i\beta}\;,\label{spin_rep}
\end{equation}
where $\sigma^\mu$ with $\mu=x,y,z$ denotes the Pauli matrices. While this representation is valid for both bosonic and fermionic partons, we will use a fermionic description in the following. This is because the experimental neutron data points towards a gapless quantum spin liquid which can only be described by fermionic spinons. Gapless bosonic spinons, in contrast, would inevitably condense which would yield a magnetically ordered state. Fermionic spinons are also in agreement with the measured linear heat capacity. Note that Eq.~(\ref{spin_rep}) only represents a valid description of the spin operator in the subspace of single fermion occupancy on each site, i.e. for $n_i\equiv f_{i\uparrow}^\dagger f_{i\uparrow}+f_{i\downarrow}^\dagger f_{i\downarrow}=1$. Most importantly, in a quantum spin liquid there cannot be long-range confining forces between spinons which would bind them into conventional $\Delta S=1$ magnetic excitations (such as spin waves of a classical magnet). Hence, in a first simple approach, the dynamics of spinons may be described by a general model of free fermions\cite{wen91,wen02}
\begin{equation}
H=\sum_{ij}\left(t_{ij}f_{i\alpha}^\dagger f_{j\alpha}+\Delta_{ij}f_{i\uparrow}f_{j\downarrow}+\text{H.c.}\right)+\mu\sum_{i}n_i\;.\label{ham_free}
\end{equation} 
Due to the experimental observation that Ca$_{10}$Cr$_7$O$_{28}$ shows only isotropic magnetic response,\cite{balz16} we will restrict to spin-independent (i.e. spin isotropic) hopping $t_{ij}$ and singlet pairing $\Delta_{ij}$. On the level of such mean-field-like quadratic Hamiltonians, the exact constraint is usually replaced by the simpler averaged constraint $\langle n_i\rangle=1$ which is equivalent to half-filled spinon bands. It is crucial to emphasize that Eq.~(\ref{ham_free}) does not yet fully describe a quantum spin liquid. The key missing ingredient which turns a simple model of free fermions into a low-energy effective theory of a quantum spin liquid are gauge degrees of freedom which correspond to phase fluctuations in the hopping and pairing amplitudes $t_{ij}$, $\Delta_{ij}$ giving rise to additional {\it spinless} excitations called visons or fluxes.\cite{wegner71,kogut79,senthil00,read89,kivelson89,punk14} Furthermore, promoting the chemical potential $\mu$ to a fluctuating field (which acts as the time-component of a gauge field), allows to fulfill the constraint $n_i=1$ exactly instead of just implementing it on average. Unfortunately, the coupling of the fermions to emergent gauge fields results in a complicated many-body theory which (at least in the generic case) cannot be easily treated. There are, however, experimental indications that the gauge excitations in Ca$_{10}$Cr$_7$O$_{28}$ are gapped while the spinon excitations are gapless (see Sec.~\ref{low_e}). In this so-called $\mathds{Z}_2$ spin-liquid scenario\cite{wegner71,kogut79,senthil00} the gauge fluctuations take the simplest possible form and only amount to variations in the sign of the hopping/pairing amplitudes (as described by the replacement $t_{ij}\rightarrow \sigma_{ij}t_{ij}$, $\Delta_{ij}\rightarrow \sigma_{ij}\Delta_{ij}$ with the gauge field $\sigma_{ij}=\pm1$). As a consequence, the coupling of the spinons to gapped vison excitations would not modify the fermionic theory in Eq.~(\ref{ham_free}) at low energies such that this simple model would still be qualitatively correct (gapped visons can at most induce short-range interactions between spinons). For these reasons, we will mostly neglect the effect of visons in our considerations and only qualitatively discuss their possible impact at the end of Sec.~\ref{comparison}.

With these arguments, the remaining theoretical task amounts to identifying a model of free spinons [such as Eq.~(\ref{ham_free})] which correctly reproduces the experimentally measured spin structure factor and the specific heat presented in Sec.~\ref{exp_results}. As explained in more detail below, the spin structure factor is given (up to weight factors) by the two-spinon spectrum of Eq.~(\ref{ham_free}). While it is rather straightforward to diagonalize Eq.~(\ref{ham_free}) with given amplitudes $t_{ij}$, $\Delta_{ij}$, $\mu$ and calculate the corresponding two-spinon excitations, the reverse, i.e., deducing a free fermion model from the spin structure factor is quite challenging. In previous works attempting a similar fitting for other compounds (see e.g. Refs.~\onlinecite{dodds13,bieri15,halimeh16,shen16,huang17,messio17,fak17}) the lattice structures and magnetic couplings were often comparatively simple and it was sometimes even sufficient to assume a single spatially uniform spinon hopping/pairing amplitude.\cite{shen16} In the case of Ca$_{10}$Cr$_7$O$_{28}$, however, there are many inequivalent ferromagnetic and antiferromagnetic nearest neighbor bonds on which the hopping and pairing amplitudes $t_{ij}$, $\Delta_{ij}$ may all be different. Hence, the most direct approach of simply testing a large number of free fermion models and searching for agreement between the measured and calculated spin structure factor represents a rather cumbersome task. Furthermore, since there is, in general, no simple map between the exchange couplings $J_{ij}$ of the original spin Hamiltonian and the amplitudes $t_{ij}$ and $\Delta_{ij}$ in a quantum spin liquid, it is also possible that the pairings/hoppings are longer-range than the interactions $J_{ij}$. Another complication arises due to the inherent gauge freedom of a parton theory which allows for a so-called {\it projective} implementation of symmetries. As a consequence of these gauge properties, the free fermion model does not need to obey all spatial symmetries of the original spin Hamiltonian, leading to an even wider class of allowed hopping and pairing amplitudes (a classification of all possible free spinon models is achieved within the {\it projective symmetry group} approach\cite{wen02}). In total, this results in a fitting problem with a large number of free parameters and a complicated map between such parameters and the target function (i.e. the spin structure factor).

Here, we try to avoid the aforementioned complications by not attempting to systematically explore all possible parameter settings for $t_{ij}$ and $\Delta_{ij}$. Rather, we will show below that based on physical arguments and experimental insights it is possible to construct a relatively simple and general spinon model which reproduces the key features of the measured spin structure factor and specific heat.

\subsection{Effective spinon model for Ca$_{10}$Cr$_7$O$_{28}$}\label{cacro}
In the first step of developing a microscopic spinon model for Ca$_{10}$Cr$_7$O$_{28}$ we will, for simplicity, neglect all pairing terms $\Delta_{ij}$. This results in a model for a so-called $U(1)$ spin liquid which is characterized by the fact that their low energy effective theory in Eq.~(\ref{ham_free}) is invariant under gauge transformations $f_{i\alpha}\rightarrow e^{i\varphi}f_{i\alpha}$ where $e^{i\varphi}$ is a complex U(1) phase. In Sec.~\ref{low_e}, we will explain how the pairings need to be reintroduced to obtain the best agreement with experimental results. Such pairings may turn the U(1) spin liquid into a $\mathds{Z}_2$ spin liquid which guarantees that the flux excitations are gapped.

An important piece of information about Ca$_{10}$Cr$_7$O$_{28}$ is that the strongest couplings in its microscopic spin Hamiltonian are ferromagnetic and act within triangular units. Since these ferromagnetic couplings are considerably larger than the antiferromagnetic ones the three spins on such triangles add up equally to form (approximate) spin-3/2 entities which represent the system's effective magnetic degrees of freedom at low energies. To account for this in our parton picture we likewise assume that the three spinon operators on a ferromagnetically coupled triangle symmetrically combine into an effective low-energy fermionic degree of freedom $c_{\text{s}I\alpha}$ via 
\begin{equation}
c_{\text{s}I\alpha}=\frac{1}{\sqrt{3}}(f_{I1\alpha}+f_{I2\alpha}+f_{I3\alpha})\;.\label{low_e}
\end{equation}
Here we have introduced a new notation which replaces the site label $i$ by two indices $I$ and $\kappa$ where $I$ denotes the ferromagnetically coupled triangle the site $i$ belongs to and $\kappa=1,2,3$ labels the sites within this triangle, i.e. we replace $f_{i\alpha}\rightarrow f_{I\kappa\alpha}$. Furthermore, the index ``s'' in $c_{\text{s}I\alpha}$ stands for the symmetric combination of spinon operators on triangle $I$. Apart from these low energy degrees of freedom there are two more linear independent spinon combinations which model energetically higher spin-1/2 excitations where the three spinons on triangle $I$ no longer equally add up. We label these combinations by $\text{a}_1$ and $\text{a}_2$ and define them by
\begin{align}
c_{\text{a}_1I\alpha}&=\frac{1}{\sqrt{2}}(f_{I1\alpha}-f_{I2\alpha})\;,\notag\\
c_{\text{a}_2I\alpha}&=\frac{1}{\sqrt{6}}(f_{I1\alpha}+f_{I2\alpha}-2f_{I3\alpha})\;.\label{high_e}
\end{align}
For the following considerations it will be convenient to illustrate the ferromagnetically coupled triangles of the bilayer kagome system by projecting them into one plane effectively resulting in a decorated honeycomb lattice, shown in Fig.~\ref{fig:bilayer}(b). In this way of drawing the lattice, the bonds between nearest neighbor triangles (which belong to different planes) are the ones with ferromagnetic interlayer couplings (red bonds) and second neighbor triangles (which lie in the same plane) are connected by antiferromagnetic intralayer couplings not shown in Fig.~\ref{fig:bilayer}(b). Also note that the unit cell consists of six sites formed of two nearest neighbor ferromagnetic triangles. Below, we will label the six sites of a unit cell by an index $\tilde{\kappa}$ where the tilde distinguishes it from the index $\kappa$ which runs only over the three sites of one ferromagnetic triangle.

The central conceptual step of the following considerations is to formulate our spinon model in terms of the effective degrees of freedom $c_{\text{s}I\alpha}$, $c_{\text{a}_1I\alpha}$, $c_{\text{a}_2I\alpha}$ instead of the original operators $f_{I\kappa\alpha}$. A generic fermionic hopping model in these degrees of freedom with hopping terms ranging up to second neighbor triangles can then be written as
\begin{align}
H&=\sum_{\langle IJ \rangle,\alpha} \left[t_{1\text{s}} c^\dagger_{\text{s}I\alpha} c_{\text{s}J\alpha}+t_{1\text{a}}\left(c^\dagger_{\text{a}_1I\alpha} c_{\text{a}_1J\alpha}+c^\dagger_{\text{a}_2I\alpha} c_{\text{a}_2J\alpha}\right)\right]\notag\\
&+\sum_{\langle\langle IJ \rangle\rangle,\alpha} \left[t_{2\text{s}} c^\dagger_{\text{s}I\alpha} c_{\text{s}J\alpha}+t_{2\text{a}}\left(c^\dagger_{\text{a}_1I\alpha} c_{\text{a}_1J\alpha}+c^\dagger_{\text{a}_2I\alpha} c_{\text{a}_2J\alpha}\right)\right]\notag\\
&+\text{H.c.}\notag\\
&+\sum_{I\alpha}\left[ \mu_\text{s} c^\dagger_{\text{s}I\alpha}c_{\text{s}I\alpha}+\mu_{\text{a}}\left(c^\dagger_{\text{a}_1I\alpha} c_{\text{a}_1I\alpha}+c^\dagger_{\text{a}_2I\alpha} c_{\text{a}_2I\alpha}\right)\right]\;.\label{ham_c}
\end{align}
Here, $\langle IJ \rangle$ ($\langle\langle IJ \rangle\rangle$) denote nearest (second) neighbor ferromagnetic triangles, i.e., $t_{1\text{s}}$ and $t_{1\text{a}}$ ($t_{2\text{s}}$ and $t_{2\text{a}}$) are nearest (second) neighbor hopping amplitudes for the low energy symmetric and high energy asymmetric spinon degrees of freedom, respectively. Furthermore, $\mu_\text{s}$ and $\mu_\text{a}$ are chemical potentials acting on the two types of spinons. We emphasize that since all three sites in a ferromagnetic triangle are symmetry-equivalent the Hamiltonian must be invariant under the exchange of indices $\text{a}_1\leftrightarrow\text{a}_2$. In this case, the Hamiltonian does not depend on the precise definition of the high-energy degrees of freedom in Eq.~(\ref{high_e}) as long as the $c$ operators all correspond to orthogonal states.

A central requirement for the model in Eq.~(\ref{ham_c}) is that $\mu_{\text{a}}$ is large enough to ensure a clear separation of low-energy and high-energy spinon modes as is visible in the measured spin structure factor showing two energy intervals with strong signal and a relatively weak response in between. As will become clear in the next subsection, under this condition the model in Eq.~(\ref{ham_c}) already reproduces some key aspects of the experimental neutron data {\it irrespective} of the precise choice of the parameters $t_{1/2\text{s}/\text{a}}$ and $\mu_{\text{s}/\text{a}}$. 

\subsection{General weight distribution of the dynamical spin structure factor}\label{general_weight_dist}
We start discussing the dynamical spin-structure factor which is defined by
\begin{align}
\mathcal{S}({\mathbf q,E})&\equiv\mathcal{S}^{zz}({\mathbf q,E})\notag\\
&=\frac{1}{N}\int_{-\infty}^{\infty} dt\sum_{ij}e^{iE t}e^{i{\mathbf q}({\mathbf r}_i-{\mathbf r}_j)}\langle S_i^z(t)S_j^z(0)\rangle\label{s_def}
\end{align}
where $N$ is the total number of sites and ${\mathbf r}_i$ denotes the position of site $i$ (because of the system's spin-isotropy we have $\mathcal{S}({\mathbf q,E})=\mathcal{S}^{xx}({\mathbf q,E})=\mathcal{S}^{yy}({\mathbf q,E})=\mathcal{S}^{zz}({\mathbf q,E})$). Particularly, we analyze some general properties of the weight distribution of $\mathcal{S}({\mathbf q,E})$ in energy and momentum space for the spinon model in Eq.~(\ref{ham_c}) and show that it qualitatively matches the experimental results even without fine-tuning of the hopping parameters.
\begin{figure*}[t]
\centering
\includegraphics[width=0.9\linewidth]{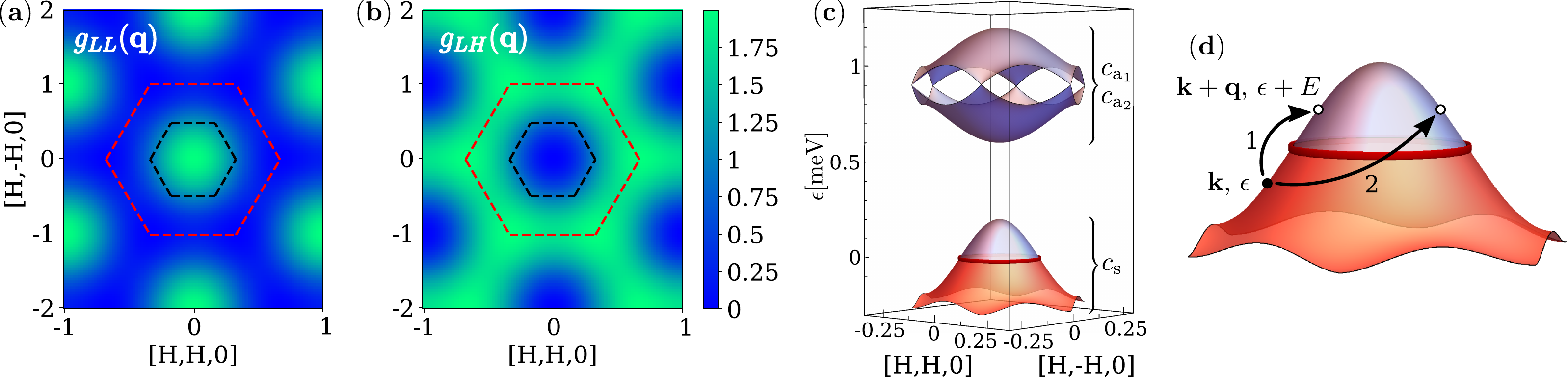}
\caption{(a), (b) Weight factors $g_{\text{LL}}({\mathbf q})$ and $g_{\text{LH}}({\mathbf q})$ of the dynamical spin structure factor $\mathcal{S}({\mathbf q,E})$ as defined in Eqs.~(\ref{s_parton})-(\ref{glh}) for the spinon hopping amplitudes given in Eq.~(\ref{par}). Here, $g_{\text{LL}}({\mathbf q})$ ($g_{\text{LH}}({\mathbf q})$) is the total weight factor for all particle-hole excitation processes within the low-energy bands formed by $c_{\text{s}}$ (between the low-energy bands formed by $c_{\text{s}}$ and the high-energy bands resulting from $c_{\text{a}_1}$, $c_{\text{a}_2}$). Black (red) dashed lines indicate the boundaries of the first (extended) Brillouin zone. (c) Spinon band structure for the same set of spinon hopping amplitudes [see Eq.~(\ref{par})]. Low energy (high energy) bands are marked by the corresponding spinon operators $c_{\text{s}}$ ($c_{\text{a}_1}$ and $c_{\text{a}_2}$) they result from. The red line marks the Fermi surface. Note that all bands are doubly degenerate. (d) Illustration of two different particle-hole excitations around the Fermi surface with a given energy $E$. Process 1 shows an excitation from an occupied state (full black dot) with momentum and energy ${\mathbf k}$, $\epsilon$ to an unoccupied state (open dot) with ${\mathbf k}+{\mathbf q}$, $\epsilon+E$ for the minimal momentum transfer ${\mathbf q}$ which linearly depends on the energy $E$. Process 2 is an example for a particle-hole excitation with larger momentum transfer.}
\label{fig:weight}
\end{figure*}

To explicitly calculate the dynamical spin-structure factor for an effective parton model, we need its eigenenergies (i.e. free spinon-band dispersion) which for Eq.~(\ref{ham_c}) we denote by $\epsilon_a({\mathbf q})$. Here, $a$ is a band index with values $a=1,2,\ldots,6$ due to the six-atomic unit cell. Note that $a=1,2$ labels the low-energy bands resulting from $c_{\text{s}I\alpha}$ while $a=3,4,5,6$ corresponds to the high-energy bands due to $c_{\text{a}_1I\alpha}$ and $c_{\text{a}_2I\alpha}$. Inserting the spin representation of Eq.~(\ref{spin_rep}) into Eq.~(\ref{s_def}) and expanding the expectation value of the fermionic operators one finds
\begin{align}
\mathcal{S}({\mathbf q,E})&=\frac{\pi}{24}\sum_{a,b}\int\frac{d^2 k}{(2\pi)^2}f({\mathbf k},{\mathbf q},a,b)[n_a({\mathbf k})-n_b({\mathbf k}+{\mathbf q})]\notag\\
&\times\delta(\epsilon_b({\mathbf k}+{\mathbf q})-\epsilon_a({\mathbf k})-E)\label{s_parton}
\end{align}
where $f({\mathbf k},{\mathbf q},a,b)$ is a weight function to be discussed further below and $n_a({\mathbf k})$ is the occupation number of an eigenstate of Eq.~(\ref{ham_c}) with band index $a$ and wave vector ${\mathbf k}$. From Eq.~(\ref{s_parton}), the spin structure factor can be interpreted as a spinon particle-hole excitation spectrum taking into account all processes where a fermion in the occupied state with energy $\epsilon_a({\mathbf k})$ is destroyed and a fermion in the unoccupied state $\epsilon_b({\mathbf k}+{\mathbf q})$ is created, leading to a contribution to $\mathcal{S}({\mathbf q,E})$ at the corresponding momentum and energy transfers ${\mathbf q}$ and $E=\epsilon_b({\mathbf k}+{\mathbf q})-\epsilon_a({\mathbf k})$, respectively. The contributions from such processes are modulated with the weight function $f({\mathbf k},{\mathbf q},a,b)$ given by
\begin{equation}
f({\mathbf k},{\mathbf q},a,b)=\left|\sum_{\tilde{\kappa}} \phi_{a\tilde\kappa}^*({\mathbf k})\phi_{b\tilde\kappa}({\mathbf k}+{\mathbf q})e^{i{\mathbf q}{\mathbf R}_{\tilde{\kappa}}}\right|^2\;, \label{weight}
\end{equation}
where $\phi_{a\tilde\kappa}({\mathbf k})$ is the eigenstate of Eq.~(\ref{ham_c}) at sublattice site $\tilde\kappa$,  wave-vector ${\mathbf k}$, and band index $a$ Furthermore, ${\mathbf R}_{\tilde\kappa}$ is the position of the sublattice $\tilde\kappa$ within a unit cell, i.e., relative to a fixed base point inside each unit cell. Since $f$ contains the overlap of the two wave functions involved in the spinon particle-hole process, which can range from zero to one, it has a significant effect on the qualitative form of the spin structure factor. To quantify this effect, we discuss the momentum integrated weight function
\begin{equation}
g({\mathbf q},a,b)=\int\frac{d^2 k}{(2\pi)^2}f({\mathbf k},{\mathbf q},a,b)
\end{equation}
to identify regions in ${\mathbf q}$ where the modulation due to $f$ enhances or suppresses the spin structure factor. Particularly, we consider the weight function $g_{\text{LL}}$ for all particle-hole excitations between the two low-energy bands
\begin{equation}
g_{\text{LL}}({\mathbf q})=\sum_{a,b=1,2} g({\mathbf q},a,b)\label{gll}\;.
\end{equation}
Note that this function only has an effect on the spin structure factor if such particle-hole processes exist in the first place, i.e. if the Fermi energy lies within these low-energy bands. The latter assumption will turn out to be an important property for the following considerations. Furthermore, we investigate the weight function $g_{\text{LH}}$ for all particle-hole excitations between the low and high-energy bands
\begin{equation}
g_{\text{LH}}({\mathbf q})=\sum_{a=1,2}\sum_{b=3,4,5,6} g({\mathbf q},a,b)\;.\label{glh}
\end{equation}
Since we assume that the high-energy bands are unoccupied in the ground state, particle-hole processes within these bands do not need to be considered. Due to the orthogonality of eigenmodes, it is clear from Eq.~(\ref{weight}) that $f({\mathbf k},{\mathbf q}=0,a,b=a)=1$ and $f({\mathbf k},{\mathbf q}=0,a,b\neq a)=0$. Hence, $g_{\text{LL}}({\mathbf q})$ has its maximum at the $\Gamma$-point (${\mathbf q}=0$) while $g_{\text{LH}}({\mathbf q})=0$ for ${\mathbf q}=0$. With the properties at ${\mathbf q}=0$ fixed, a closer numerical inspection shows that even for finite ${\mathbf q}$-vectors $g_{\text{LL}}({\mathbf q})$ and $g_{\text{LH}}({\mathbf q})$ are rather insensitive to the precise values of the parameters $t_{1/2\text{s}/\text{a}}$ and $\mu_{\text{s}/\text{a}}$ in Eq.~(\ref{ham_c}). As an example, we show in Fig.~\ref{fig:weight}(a), (b) the quantities $g_{\text{LL}}({\mathbf q})$ and $g_{\text{LH}}({\mathbf q})$ for the hopping amplitudes $t_{1\text{s}}=0$, $t_{2\text{s}}=0.05$ meV, $t_{1\text{a}}=0.1$ meV, $t_{2\text{a}}=0$, $\mu_\text{s}=-0.1$ meV, $\mu_\text{a}=0.9$ meV (this parameter setting will be considered further below). Particularly, $g_{\text{LL}}({\mathbf q})$ is found to be sizeable everywhere inside the first Brillouin zone but drops off rapidly beyond its boundaries. On the other hand, $g_{\text{LH}}({\mathbf q})$ is large in significant portions of ${\mathbf q}$ space (even up to the edges of the extended Brillouin zone) except inside the first Brillouin zone.

These properties have various important consequences for the form of $\mathcal{S}({\mathbf q,E})$ obtained from the spinon model in Eq.~(\ref{ham_c}).

(i) At low energies where only bands with $a,b=1,2$ contribute in Eq.~(\ref{s_parton}), i.e., the function $g_{\text{LL}}({\mathbf q})$ determines the general weight distribution, the spin-structure factor is mainly concentrated inside or around the boundaries of the first Brillouin zone. To continue this line of argument, if the two low-energy bands never fulfill the condition $n_1({\mathbf k})=1$ and $n_2({\mathbf k})=0$ for any wave vector ${\mathbf k}$ (e.g., if the bands 1 and 2 are degenerate), a particle-hole excitation with ${\mathbf q}=0$ cannot occur which further suppresses the spin structure factor at and around the $\Gamma$-point for small $E$. Combining both properties, $\mathcal{S}({\mathbf q,E})$ inevitably shows a pattern of ring-like magnetic response at low energies which is approximately distributed along the boundaries of the first Brillouin zone. This behavior matches the experimental observation and will be demonstrated more explicitly in the next subsection. We emphasize that an important requirement for this argument to hold is that the Fermi level passes through the bands $a=1,2$. The Fermi surface which is formed with these bands is also crucial for explaining the linear heat capacity and the low-energy behavior of the spin structure factor further discussed in Sec.~\ref{comparison}.

(ii) If $\mu_\text{a}$ is chosen sufficiently large, i.e., there is a clear separation between spinon bands with $a=1,2,$ and $a=3,4,5,6$, a regime of low signal at intermediate energies $E$, as observed experimentally, can be realized.

(iii) At high energies, the more spread out weight function $g_{\text{LH}}({\mathbf q})$ allows the dynamic spin structure factor to be sizeable within large parts of reciprocal space also reaching out to the edges of the extended Brillouin zone. This behavior again matches the experimental finding.

In the next subsection we will demonstrate these properties based on a numerical evaluation of Eq.~(\ref{s_parton}) for a particular choice of spinon hopping parameters and spinon chemical potentials. 

\subsection{Comparison with the measured spin-structure factor}\label{comparison}
The aforementioned weight factors $g_{\text{LL}}({\mathbf q})$ and $g_{\text{LH}}({\mathbf q})$ largely determine the form of the dynamical spin-structure factor and allow to reproduce some of its key features for a wide range of amplitudes $t_{1/2\text{s}/\text{a}}$ and $\mu_{\text{s}/\text{a}}$ in Eq.~(\ref{ham_c}). However, the identification of the optimal set of these six parameters for which the agreement between theory and experiment is best, still represents a non-trivial task. Moreover, systematically justifying such amplitudes, i.e., developing a microscopic theory of how these parameters arise from the exchange couplings is, likewise, very difficulty and goes beyond the current understanding of quantum spin liquids. We will not try to pursue these directions here. Rather, we will show that for the particular set of amplitudes given by
\begin{align}
&t_{1\text{s}}=0,\;t_{2\text{s}}=0.05\text{ meV},\;t_{1\text{a}}=0.1\text{ meV},\;t_{2\text{a}}=0,\notag\\
&\mu_\text{s}=-0.1\text{ meV},\;\mu_\text{a}=0.9\text{ meV}\label{par}
\end{align}
the properties (i), (ii), and (iii) of Sec. \ref{general_weight_dist} are fulfilled, leading to an approximate agreement with experimental observations. An illustration of the corresponding spinon dispersion is plotted in Fig.~\ref{fig:weight}(c). We wish to emphasize again that we do not claim that these exact values are realized in Ca$_{10}$Cr$_7$O$_{28}$ since there are wide parameter regions which yield similar (or possibly even better) agreement. Our aim here is to demonstrate that based on the considerations leadings to Eq.~(\ref{ham_c}) the overall form of the measured spin structure factor can be explained rather straightforwardly in an effective spinon picture. While the amplitudes in Eq.~(\ref{par}) do not result from a systematic optimization they can still be motivated based on various physical arguments and observations:
\begin{figure}[t]
\centering
\includegraphics[width=0.99\linewidth]{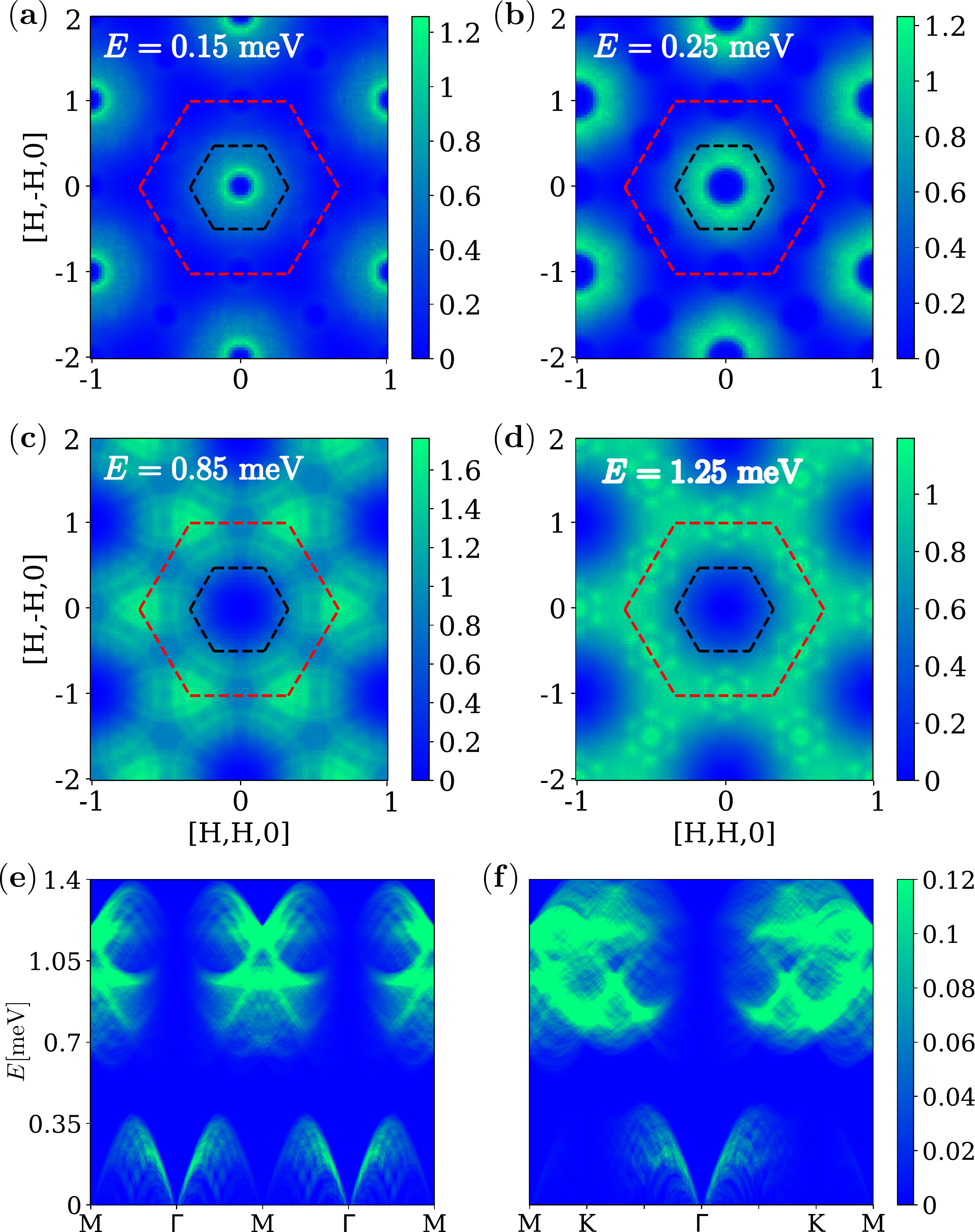}
\caption{Calculated spin-structure factor for the effective spinon model in Eq.~(\ref{ham_c}) using the representation of $\mathcal{S}({\mathbf q,E})$ from Eq.~(\ref{s_parton}) and the spinon parameters in Eq.~(\ref{par}). The plots (a)-(d) show the spin structure factor in momentum space for the same fixed energies $E$ as for the experimental neutron data in Fig.~\ref{fig:neutron} (a)-(d). Black (red) dashed lines indicate the boundaries of the first (extended) Brillouin zone. The figures (e) and (f) show $\mathcal{S}({\mathbf q,E})$ as a function of energy along two momentum space directions to compare with Fig.~\ref{fig:neutron} (e) and (f), respectively. The data in (a)-(d) has been convoluted with a gaussian distribution function to match the experimental resolution while in (e) and (f) the finite energy resolution and perpendicular ${\mathbf q}$-integration have not been taken into account. Note that the magnetic form factor of the Cr$^{5+}$ ions is not included in these plots.}
\label{fig:s_theory}
\end{figure}

(a) As already mentioned in point (i) of Sec. \ref{general_weight_dist} the observed small signal in $\mathcal{S}({\mathbf q,E})$ at the $\Gamma$-point for small energies can be guaranteed by requiring that the band occupations $n_1({\mathbf k})=1$ and $n_2({\mathbf k})=0$ are never realized for any ${\mathbf k}$. This property can be most straightforwardly fulfilled if the bands $a=1$ and $2$ are degenerate which is realized when setting $t_{1\text{s}}=0$ and $t_{2\text{s}}\neq0$ (in this situation the two sublattices of the effective honeycomb lattice of ferromagnetic triangles decouple leading to two identical low-energy bands).

(b) In the experimental spin-structure factor the two energy regions of strong signal, i.e. around $0.3$ meV and around $1$ meV are similar in magnitude. In our spinon model, however, the low-energy region arises from particle-hole excitations within the spinon bands with $a=1,2$ while the high-energy response results from excitations between bands with $a=1,2$ and $a=3,4,5,6$. Hence, the latter process involves more spinon bands which typically leads to more possibilities of particle-hole excitations and, therefore, to higher intensities of $\mathcal{S}({\mathbf q,E})$ at large energies as compared to small energies. This particularly occurs when the spinon bands with $a=3,4,5,6$ are degenerate (or energetically nearby) such that two-particle processes with the same ${\mathbf q}$ and $E$ but different $a=3,4,5,6$ all add up. To avoid such effects, we make sure that these bands are well separated from each other which is realized when setting $t_{1\text{a}}\neq0$ and $t_{2\text{a}}=0$ leading to a spinon Dirac cone dispersion in the higher bands. (We note, however, that since $c_{\text{a}_1}$ and $c_{\text{a}_2}$ need to appear symmetrically in Eq.~(\ref{ham_c}) it cannot be avoided that pairs of high-energy spinon bands with $a=3,4$ and $a=5,6$, respectively, are always degenerate).  

(c) The size of the remaining non-vanishing spinon parameters $t_{2\text{s}}$, $t_{1\text{a}}$, $\mu_\text{s}$, and $\mu_\text{a}$ are adjusted such that the extent of the two high-intensity regions of $\mathcal{S}({\mathbf q,E})$ as well as the size of apparent gapped region between them comes out approximately correct.

We start discussing the low-energy region of the calculated spin structure factor [Fig.~\ref{fig:s_theory}(a), (b) for $E=0.15$ meV and $E=0.25$ meV, respectively] and compare it with the measured data [Fig.~\ref{fig:neutron}(a), (b) for the same energies]. As can be seen, the combined effects of the weight function $g_{\text{LL}}({\mathbf q})$ and the suppressed response around the $\Gamma$ point due to the argument given in Sec.~\ref{general_weight_dist} (i) lead to a ring-like pattern in $\mathcal{S}({\mathbf q,E})$. Furthermore, as is evident from the ${\mathbf q}$ space cuts along two momentum directions in Fig.~\ref{fig:s_theory}(e), (f) the diameter of the rings increases linearly with energy and, hence, the calculated spin structure factor exhibits a similar characteristic V-shape as the experimental data in Fig.~\ref{fig:neutron}(e), (f). This linear onset of response can be directly explained by the presence of a spinon Fermi surface in our parton model around which the spinons have an approximate linear dispersion. As a consequence, any particle-hole excitation near the Fermi surface with a given energy $E$ (i.e., where a spinon from an occupied state with momentum ${\mathbf k}$ and energy $\epsilon$ changes into an unoccupied state with momentum ${\mathbf k}+{\mathbf q}$ and energy $\epsilon+E$) requires at least a momentum transfer $|{\mathbf q}|$ which linearly depends on the energy, i.e., $E=v_\text{sp}|{\mathbf q}|$ where $v_\text{sp}$ is the spinon Fermi velocity. For an illustration of such particle-hole excitations, see Fig.~\ref{fig:weight} (d). We, hence, suggest that the V-shaped scattering signal at low ${\mathbf q}$ and $E$ is a direct consequence of a spinon Fermi surface.

Moving up in energy, the dynamical spin-structure factor exhibits a regime of relatively weak signal at around $0.6$ meV followed by stronger response which reaches up to approximately $1.4$ meV, see Fig.~\ref{fig:s_theory} (e) and (f). As shown in Fig.~\ref{fig:s_theory} (c) and (d), in this latter regime, the strong signal extends over large parts of reciprocal space (except for small ${\mathbf q}$) and, particularly, fills the area between the first and the extended Brillouin zone. Note that our calculated spin-structure factor in this high-energy region shows a rather complex pattern of intensity which varies on smaller momentum and energy scales [see Fig.~\ref{fig:s_theory} (c)-(f)] and which sensitively depends on the precise choice of parameters in the effective spinon model. A detailed comparison of such features with the measured spin-structure factor data would possibly allow for more insights into the spinon band structure, however, given the uncertainties of these parameters and the limited experimental resolution we only discuss the more extended features here.

While our calculated spin-structure factor reproduces some overall characteristic features of the measured data, pronounced differences are also revealed. Most obviously, the measured signal spreads out into much larger regions in reciprocal space than the calculated response. For example, the experimental spin-structure factor along the momentum direction $M$-$K$-$\Gamma$-$K$-$M$ and at small energies $E$ [see Fig.~\ref{fig:neutron} (f)] remains large throughout this line, while our spinon model predicts a rapidly decaying signal for increasing $|{\mathbf q}|$ [see Fig.~\ref{fig:s_theory} (f)]. Such differences are also pronounced in the high-energy region where the measured response shows a rather uniform distribution in the entire ${\mathbf q}$-space, even inside the first Brillouin zone. In contrast, our calculations always reveal a vanishing spin-structure factor at high energies around the $\Gamma$-point. Since this latter suppression of signal is a consequence of the corresponding particle-hole excitations involving two orthonormal spinon states [see the weight factor $f({\mathbf k},{\mathbf q}=0,a,b\neq a)$ in Eq.~(\ref{weight})] it is an exact property that cannot be avoided within our free spinon model of Eq.~(\ref{ham_c}). A finite spin-structure factor in the high-energy regime at ${\mathbf q}=0$ must therefore have an origin outside this model. Indeed, as already explained in Sec.~\ref{partons}, the Hamiltonian in Eq.~(\ref{ham_c}) misses gauge field degrees of freedom which manifest as phase fluctuations of all amplitudes $t_{1/2\text{s}/\text{a}}$, $\mu_{\text{s}/\text{a}}$. Such degrees of freedom (called visons or fluxes) represent an essential ingredient of effective theories for quantum spin liquids. While the full theory consisting of fermionic partons coupling to emergent gauge fields cannot be easily solved, special cases still allow for a closer investigation, see for example Ref.~\onlinecite{punk14} studying the effects of gauge fluctuations in a purely antiferromagnetic monolayer kagome Heisenberg system. Such works indicate that gauge fluctuations lead to a smearing of the spin-structure factor which, hence, becomes more spread-out in momentum space. We, therefore, propose that the inclusion of gauge fluctuations in our model might yield a better agreement between the calculated and measured spin-structure factor. Such an analysis, however, goes beyond the scope of the present work.

\subsection{Heat capacity and spinon pairing}\label{low_e}
The above approach of modeling the measured spin-structure factor naturally leads to a spinon hopping model with a Fermi surface. A spinon Fermi surface is again consistent with the linear heat capacity at low temperatures. Yet, the picture developed so far is problematic due to various reasons:

(I) When introducing gauge fields in our pure fermionic hopping model, the resulting theory has a U(1) gauge structure implying that the gauge excitations are gapless.\cite{savary17,hickey19} These low-energy excitations would give an additional contribution to the heat capacity at low temperatures changing the linear behavior to a $T^{2/3}$ dependence\cite{motrunich05} which is not observed experimentally. At the lowest accessible temperatures (i.e., $T<0.1$ K) the measured data rather seems to have a $T^2$ behavior, see the inset of Fig.~\ref{fig:heat_capacity} (b), showing the heat capacity in a double-logarithmic plot .

(II) The stability of U(1) quantum spin liquids in two dimensions is generally questionable since the underlying U(1) gauge theory has been argued to be unstable with respect to monopole proliferation.\cite{polyakov77,hermele04,rantner02,herbut03,herbut03_2} This effect may turn a quantum spin liquid into a trivial (e.g. magnetically ordered) state.

(III) As explained in Sec.~\ref{partons}, a free spinon model as shown in Eq.~(\ref{ham_free}) is subject to the constraint of half filling, i.e., $\langle n_i\rangle=1$. However, since the Fermi level cuts through the lowest two of six bands, our model is less than half-filled.

Fortunately, all three problems may be simultaneously solved by introducing spinon pairing terms $\Delta_{ij}$ in Eq.~(\ref{ham_free}). Firstly, spinon pairing breaks down the gauge structure from U(1) to $\mathds{Z}_2$. This gaps out the flux excitations which therefore do not contribute to the heat capacity at sufficiently low temperatures. Gapped flux modes also guarantee the stability of a $\mathds{Z}_2$ gauge theory such that a free parton model may be used as a qualitatively correct description of the spinful excitations of a quantum spin liquid.\cite{wen91} Finally, a spinon model with finite pairing may fulfill the parton constraint even if the corresponding model with all $\Delta_{ij}$ set to zero is not half filled. To be more precise, the generalized constraint in the presence of pairing may be formulated as $\partial E_\text{ground}/\partial \mu=0$ where $E_\text{ground}$ is the ground state energy of Eq.~(\ref{ham_free}).\cite{wen02,note1}

These arguments indicate that pairing is a necessary ingredient in our effective spinon theory. We will now demonstrate that its inclusion also allows for a more accurate modeling of the heat capacity at low temperatures. It is important to emphasize, however, that a finite spinon pairing does not imply that the system becomes a real superconductor, since spinons do not carry charge. In the following, we take the measured heat-capacity data literally and strictly assume that it is of pure spinon origin in the temperature regime up to $\approx 0.5$ K (i.e., we exclude possible contributions from phonons or flux excitations and also neglect impurity scattering). We first reiterate that under this assumption an intact Fermi surface would lead to a perfect linear heat capacity. While its measured behavior is indeed mostly linear in the temperature regime $T\lesssim 0.5$ K, a small reduction from linearity is observed for $T\lesssim 0.1$ K, see Fig.~\ref{fig:heat_capacity} (b). This can be interpreted as a signature of pairing in the low energy spinon band which gaps out the Fermi surface (at least partially) and, as a consequence, reduces the density of states and heat capacity at energies/temperatures below the gap. If chosen sufficiently small, spinon pairing could, hence, explain this low-temperature deviation from linearity but at the same time keep the almost perfect linear behavior at $0.1$ K $\lesssim T\lesssim0.5$ K intact (a similar scenario for triangular lattice compounds is discussed in Ref.~\onlinecite{grover10}). As we will see below, the corresponding pairing amplitudes are on an energy scale much smaller than the minimal energy $\approx 0.15$ meV down to which the spin-structure factor is measured. The inclusion of pairing, therefore, has a negligible effect on our results in the last subsection and does not affect the conclusions already made.

Given that the spinon hopping amplitudes discussed in Sec.~\ref{comparison} are already subject to large uncertainties we will not attempt here to determine explicit pairing parameters $\Delta_{ij}$ and to fine-tune them according to the constraint $\partial E_\text{ground}/\partial \mu=0$. Rather, we will concentrate on the low energy spinon band and discuss the general momentum space dependence of the pairing gap which yields the best agreement with the measured data. Two pairing scenarios seem possible: Firstly, pairing may be of $s$-wave type [i.e., with a constant gap function $\Delta({\mathbf k})\equiv\Delta_s$ in reciprocal space] gapping out the entire Fermi surface which leads to an activated behavior in the heat capacity, i.e. $C(T)\propto \exp[-\Delta_s/(k_\text{B}T)]$ at the lowest temperatures $T\lesssim \Delta_s$. Secondly, pairing may gap out the Fermi surface except for discrete nodal points. In this case, the three-fold in-plane rotation symmetry of the system suggests $f$-wave pairing, which is characterized by six gapless Dirac points, see Fig.~\ref{fig:heat_capacity} (c). An $f$-wave pairing gap has the momentum structure
\begin{equation}
\Delta({\mathbf k})=\Delta_f|\sin(3\varphi_{\mathbf k}+\varphi_0)|\;,\label{f_wave}
\end{equation}
where $\varphi_{\mathbf k}$ is the polar angle in momentum space (for cartesian coordinates) with $\tan(\varphi_{\mathbf k})=k_y/k_x$ and $\varphi_0$ is a possible constant offset. In Fig.~\ref{fig:heat_capacity} (b), we compare the measured data with the best fits for both cases [making use of Eq.~(\ref{f_wave}) with $\Delta_f=0.039$ meV in the case of $f$-wave pairing] and also show the strictly linear behavior for an intact Fermi surface, i.e., without any pairing (black line). As can be seen, the $f$-wave pairing scenario yields the best agreement with the experimental data and, therefore, seems most reasonable under the above assumptions. The resulting $f$-wave low-energy spinon bands are depicted in Fig.~\ref{fig:heat_capacity} (c). Note that the Dirac cones yield a quadratic behavior of the heat capacity at the lowest temperatures $T\lesssim \Delta_f$ in agreement with the experimental data in the inset of Fig.~\ref{fig:heat_capacity} (b). The spinon dispersion can be thought of as originating from the intersection of two bands (i.e., the original spinon band from pure hopping and its particle-hole reversed version) which are gapped out along the Fermi surface with a gap according to Eq.~(\ref{f_wave}). Note that the results of our fits are independent of the diameter of the initial Fermi surface and the Fermi velocity which only enter as an overall scaling factor of the heat capacity.

We wish to conclude this analysis with a remark of caution: While our effective spinon model for Ca$_{10}$Cr$_7$O$_{28}$ explains the experimental heat capacity and spin-structure factor on a qualitative level, it relies on assumptions which seem physically well-founded but are hard to prove rigorously based on currently available experimental data. This, particularly, applies to the assumption of the pure spinon origin of the measured low-energy/low-temperature data which we hope will be further scrutinized in future experimental studies.

\section{Conclusion}\label{conclusion}
We have presented new experimental lower temperature heat capacity and highly detailed dynamical spin-structure factor data for the quantum spin liquid candidate material Ca$_{10}$Cr$_7$O$_{28}$. The measured heat capacity $C(T)$ shows an almost perfect linear temperature dependence in the range $0.1$ K $\lesssim T\lesssim0.5$ K. While this type of behavior has also been (approximately) observed in other proposed quantum spin liquid materials\cite{yamashita08,yamashita10,yamashita11} and is often interpreted as a signature of fermionic spinon excitations with a Fermi surface, in Ca$_{10}$Cr$_7$O$_{28}$ the linear dependence appears remarkably accurate and occurs within an extended temperature interval. The overall very diffuse scattering signal of the measured dynamical spin structure factor $\mathcal{S}({\mathbf q,E})$ without any well-defined spin-wave excitations further supports the existence of a quantum spin-liquid ground state with spinon excitations. Two energetically well separated bands of scattering are observed in the spin structure factor. A lower regime of magnetic response shows broad ring-like structures around the edges of the first Brillouin zone whose diameter increases with increasing energy. In the high-energy regime of magnetic scattering the intensities are located at larger momenta ${\mathbf q}$ reaching out to the edges of the extended Brillouin zone.

Guided by these observations, we model the system's fundamental spinful excitations by fermionic spinons and develop a microscopic theory for the dynamics of these quasiparticles. The key conceptual property of our model is that the three spinons on a ferromagnetically coupled triangle {\it symmetrically} combine into new effective spinon degrees of freedom (which we denote by $c_\text{s}$-operators), accounting for the strong ferromagnetic interactions on such bonds and giving rise to the low-energy regime of scattering in the spin structure factor. We further introduce two asymmetric combinations of spinons (called $c_{\text{a}_1}$ and $c_{\text{a}_2}$) to model the high-energy behavior of the spin structure factor. We show that already under rather weak assumptions, such as the existence of a spinon Fermi surface in the bands formed by $c_\text{s}$, a generic hopping model in our new spinon operators correctly describes the different patterns of scattering in the aforementioned two regimes of the dynamical spin structure factor. Moreover, the existence of a spinon Fermi surface is also crucial in explaining the observed V-shaped onset of signal in $\mathcal{S}({\mathbf q,E})$ at small energies and the linear temperature dependence of the heat capacity. The simultaneous explanation of various different experimental observations by a spinon Fermi surface, hence, provides strong evidence for such a behavior.
\begin{figure*}[t]
\centering
\includegraphics[width=0.99\linewidth]{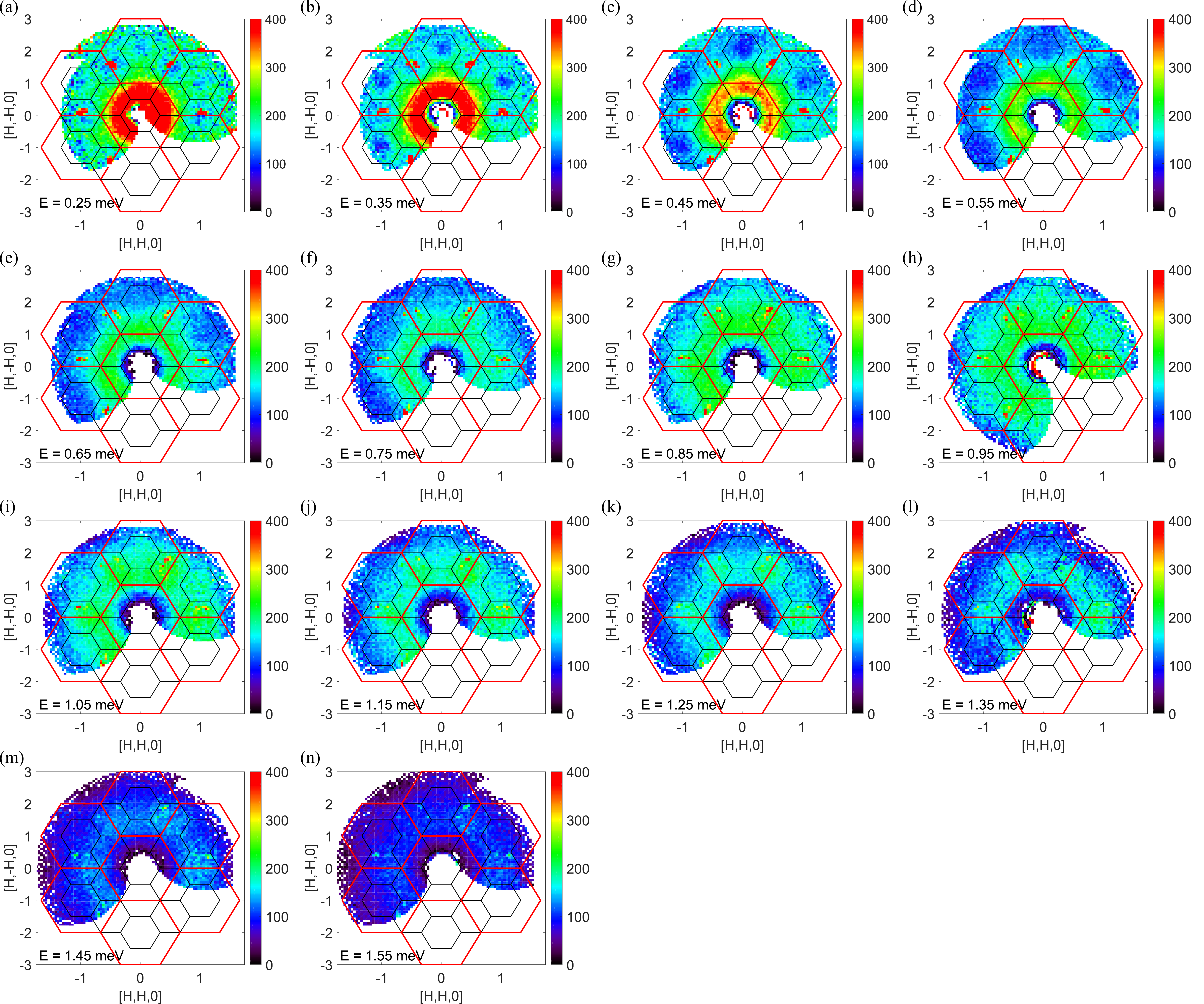}
\caption{Full inelastic neutron scattering data for Ca$_{10}$Cr$_7$O$_{28}$ at various fixed energy transfers as indicated in each subplot. See text for details.}
\label{fig:appendix}
\end{figure*}

We further provide various arguments for the need of additional weak spinon pairing terms in our spinon theory, resulting in a quantum spin liquid with an effective $\mathds{Z}_2$ gauge structure. We demonstrate that a proper choice for the spinon pairing gap allows us to correctly describe a deviation of the heat capacity from linearity at the lowest accessible temperatures. Putting together all experimental evidence and theoretical arguments, we propose a $\mathds{Z}_2$ spin liquid scenario for Ca$_{10}$Cr$_7$O$_{28}$ with an almost intact spinon Fermi surface that is only weakly gapped out by small $f$-wave spinon pairing terms, leaving behind six nodal Dirac points.

We conclude that the presented experimental data on Ca$_{10}$Cr$_7$O$_{28}$ allows for a remarkably comprehensive and coherent description of its hypothetical spin-liquid ground state, representing a rare situation compared to other spin-liquid candidate materials currently investigated. Concerning future directions of research, it will certainly be desirable to test our effective spinon theory with further experimental probes such as low-temperature susceptibility and thermal conductivity measurements. From the theory side, the precise effect of gauge excitations (visons) remains an open question which, however, goes beyond the scope of this paper. While their inclusion inevitably results in a non-trivial many-body problem, at this stage it might already be illuminating to account for their effects in a perturbative approach. We leave such investigations for future studies.

\appendix*
\section{Full neutron scattering data}
In Fig.~\ref{fig:appendix} we present the full inelastic neutron scattering dataset. The data was measured in the kagome bilayer plane ($a$-$b$ plane) within the spin liquid phase at temperatures below $T=0.1$ K. The data were collected on the MACS spectrometer at NIST and each subplot shows the scattering pattern at a different fixed energy transfer (as indicted in the plot). The intensity is indicated by the colors. The final neutron energy was fixed at $3.0$ meV and the incident neutron energy was varied to change the energy transfer. The energy resolution increases with energy transfer from $\Delta E=0.17$ meV at $E=0.25$ meV to $\Delta E=0.3$ meV at $E=1.55$ meV. The data was collected by rotating the sample over an angular range of $150^\circ$ in steps of $2^\circ.$ While the scattering angle covers a range of approximately $90^\circ$ with a step size of approximately $2.5^\circ$. A measurement of an empty sample holder was used to indicate the background and was subtracted from the data. For energy transfers $E\leq0.45$ meV this background is unreliable within the lowest wave vector extended Brillouin zone. For this reason these regions have been excluded in Fig.~\ref{fig:neutron} of the main paper.

\section*{Acknowledgements}
We thank Nic Shannon and Han Yan for fruitful discussions. C. B. acknowledges support from the U.S. Department of Energy, Office of Science, Basic Energy Sciences, Division of Scientific User Facilities. This work was partially supported by the German Research Foundation within the CRC 183 (project A02) and the SFB/TR 49. Access to MACS was provided by the Center for High Resolution Neutron Scattering, a partnership between the National Institute of Standards and Technology and the National Science Foundation under Agreement No. DMR-1508249.

\bibliography{refs}

\begin{thebibliography}{68}
\bibitem [{not()}]{note1}%
  \BibitemOpen
  \href@noop {} {}\bibinfo {note} {Note that there are two more conditions for
  the parton constraint which have the form $\partial E_\text{ground}/\partial
  \mu'=0$ and $\partial E_\text{ground}/\partial \mu''=0$. Here $\mu'$ and
  $\mu''$ are two additional Lagrange multipliers of pairing type which enter
  the Hamiltonian through the term $\sum_i
  [(\mu'+i\mu'')(f_{i\uparrow}f_{i\downarrow}-f_{i\downarrow}f_{i\uparrow})+\text{H.c.}]$.}\BibitemShut
  {Stop}%
\end{thebibliography}

\end{document}